\documentclass{article}
\usepackage[utf8]{inputenc}
\usepackage{authblk}
\usepackage{setspace}
\usepackage[margin=1.25in]{geometry}
\usepackage{graphicx}
\graphicspath{ {./figures/} }
\usepackage{subcaption}
\usepackage{amsmath}
\usepackage{lineno}

\usepackage[style=nejm, 
citestyle=numeric-comp,
sorting=none,backref=true]{biblatex}
\usepackage{hyperref}

\title{Interior Controls on the Habitability of Rocky Planets}

\author[1*]{Cedric Gillmann}

\affil[1]{ETH Zürich, Institute for Geophysics,
            Sonneggstrasse 5, 
            Zürich,
            8093, 
            Switzerland}
\affil[*]{cgillmann@ethz.ch}

\author[2,3]{Kaustubh Hakim}

\affil[2]{KU Leuven, Institute of Astronomy,
           Celestijnenlaan 200D, 
            Leuven,
            3001, 
           Belgium}

\author[1]{Diogo Louren\c{c}o}

            \affil[3]{Royal Observatory of Belgium,
           Ringlaan 3, 
           Brussels,
            1180, 
            Belgium}

\author[4]{Sascha P. Quanz}

\affil[4]{ETH Zürich, Institute for Particle Physics and Astrophysics,
            aWolfgang-Pauli-Str. 27, 
            Zürich,
           8093, 
            Switzerland}

\author[5]{Paolo A. Sossi}

\affil[5]{ETH Zürich, Institute for Geochemistry and Petrology,
            Clausiusstrasse 25,
           Zürich,
            8092,
            Switzerland}

\begin{document}

\maketitle

\begin{abstract}
No matter how fascinating and exotic other terrestrial planets are revealed to be, nothing generates more excitement than announcements regarding their habitability. From the observation of Mars to present-day efforts toward Venus and the characterization of exoplanets, the search for life, or at least environments that could accommodate life, has been a major drive for space exploration. So far, we have found no other unquestionably habitable world besides Earth. The conditions of the habitability of terrestrial planets have proved elusive, as surface conditions depend on the complex interplay of many processes throughout the evolution of a planet. Here, we review how the interior of a rocky planet can drive the evolution of surface conditions and the atmosphere. Instead of listing criteria assumed to be critical for life, we discuss how the bulk-silicate planet can affect the onset, continuation and cessation of habitability. We then consider how it can be observed and current efforts towards this end.
\end{abstract}


\section{Introduction}
\label{sec:intro}

Investigations into habitability have provided ample opportunity for reflection as to its nature, but also planetary evolution processes in general. Agreeing on a definition for habitability has proved
troublesome. 
Rather than providing a definition, we consider that the concept can be distilled to a simple question: can life develop there? Unfortunately, the notion of life is heavily biased toward life as we know it on Earth, and even the question of "development" remains vague: is survival enough? Does it include its proliferation instead? Does it cover the emergence of life? Would the requirements not change with time, and because of life itself? The interested reader could consult the review by \cite{cockell2016} for a discussion of the definition of the term.

It is nonetheless possible to draw a list of requirements that any living organism we can imagine would need to remain active. Those would include (i) a solvent (water, on Earth), (ii) energy and (iii) the presence of various elements and nutrients in sufficient abundances and forms to facilitate (i) and (ii). 
It has been argued that energy was likely easily accessible in general, from solar radiation, redox potential or other processes, and that life as we know it makes use of some of the most common building blocks (elements) in the universe, the CHONPS. Even minor elements that are used by life, such as Fe or Mg, for instance, are relatively abundant on the surface of the Earth (even if those may vary from one specific organism to the other). Finally, the solvent common to known life, water, is one of the most abundant molecules in the Solar System. 

Indeed, spectroscopic surveys show that compositions of stars in the solar neighbourhood are similar to that of our Sun \cite{hinkel2014stellar}. However, despite the ubiquity of the right elemental building blocks, the universe does not appear to be teeming with life. This indicates the need for the appropriate physico-chemical conditions to arise on planetary surfaces. For example, water would need to be liquid, which imposes further constraints on the extant pressure and temperature conditions. 

There are also important differences between conditions required for the emergence of life and its continuation: no credible model for abiogenesis has been proposed, we are still unable to produce new life through laboratory experiments and have not recorded spontaneous life formation \cite{seckbach2012genesis,stueken2013did}. 

Many specific characteristics have been proposed as essential to planetary habitability \cite[see][]{cockell2016}. Some are related a planet's galactic location, its star and orbit \cite{seager2013exoplanet}, the pattern of the other planets in the system or the presence of a large moon \cite{laskar1993stabilization}. Other are related to the planet's properties like its composition \cite{jellinek2015connections}, inherited from its origins (see section \ref{sec:onset}), its size/mass (see Section \ref{sec:conv}) or processes \cite[such as plate tectonics, for example][]{foley2015role}.

However, it has been suggested that piling up these criteria will inherently result in a rare-Earth-type situation \cite{brownlee2000rare} that only an infinitesimal fraction of terrestrial planets could fulfil \cite{lenardic2019different}. In short, precise lists describe only Earth. While it is possible that Earth could be the template for all habitable worlds, and that those are incredibly rare, limiting our approach leaves gaps in our understanding of life in general. 

In the end, based on the above, if a single criterion is to be kept, it should be the presence of liquid water. Disregarding individual organisms characteristics, in terms of biology, liquid water is deemed necessary both for the emergence and persistence of life \cite[e.g.,][]{dartnell2011}, and therefore roughly demarcates a range of surface conditions amenable to the occurrence of life. Investigating how a planet can reach and maintain conditions compatible with the presence of liquid water remains the best general-purpose criterion we can apply.

This idea was the seed for the development of the concept of habitable zone \cite[see][for example, but references to this concept date back to at least the 19$^{th}$ century]{lammer2009,lingam2021}. The circumstellar (abiotic) habitable zone \cite[HZ][]{callejas2014} is defined as the shell of space around a star where liquid water can be maintained at the surface of a planet. 
Its inner boundary is set by the maximum energy flux that can be received in the presence of liquid water (see Section \ref{sec:cess}). 
Its outer boundary is set where the maximum greenhouse can no longer sustain temperatures high enough for liquid water \cite[][]{kopparapu2013}. Most depictions of the HZ assume an Earth-like planet with an Earth-like atmosphere, highlighting the importance of atmospheric composition.

 A mix of N$_2$, CO$_2$ and H$_2$O \cite[][]{kasting1993} is often considered, but other gases would contribute, like CH$_4$ or H$_2$ \cite[e.g.][]{pierrehumbert2011}. The HZ also varies with the type of star considered and its age, as its luminosity changes throughout its evolution \cite[e.g.,][]{rugheimer2013,rugheimer2018}.

This simple definition has historically been used to select targets that are deemed likely to be habitable (see, e.g. \cite{turnbull2003target}). 

A more complete definition would consider the whole planet as a dynamic system, possibly including life itself. It remains uncertain how such extended definitions of habitable zones could be translated into practical parameters or observables, but future observation and characterization of exoplanets will 
help refine this discussion.

The presence of liquid water implies sufficiently clement conditions that H$_2$O (l) is more stable than its gaseous equivalent, water vapour. Therefore, the habitability of a terrestrial planet,  depends on the nature of its atmosphere. Hence, processes that affect the volatile inventory (such as the accretion history or atmospheric escape) or the atmosphere response to solar radiation (radiative transfer, the greenhouse effect or atmosphere dynamics) influence planetary habitability. However, as one of the drivers of long-term planetary evolution, the state of the interior of terrestrial planets is critical to the origin and evolution of the atmosphere and surface conditions. The dynamics of the mantles of planets provides a link between the atmosphere/surface and the deep interior, which extends down to the core \cite{driscoll2013,gillmann2022}. Mantle thermal history, dynamics and eventually melting cause volcanism and volatile delivery to the atmosphere. Interior dynamics also play a part in the regulation of the climate on Earth, by recycling surface material and volatiles into the interior. This cycle is deeply intertwined with outgassing, weathering and subduction. On Earth, it notably affects the CO$_2$ concentration inn the atmosphere and is named long-term carbon cycle. Likewise, the thermal evolution of the core governs the generation of a magnetic field that affects atmospheric escape processes and long-term atmospheric evolution. 

In short, a comprehensive picture of planetary evolution 
requires consideration of the planet as a complex system, that depends on the efficiency of numerous feedback mechanisms. Here, we focus on aspects relative to the interior of rocky planets and only include succinct descriptions of the interactions between those mechanisms and the interior of planets for clarity. The present review focuses on the relationship between habitability and the silicate planet. Initially, and in order to set the boundary conditions from which a given rocky planet is thought to have evolved, the molten state of the mantle is considered, before transitioning to a solid state, and its long-term interactions with the surface and atmosphere of the planet. 

Habitability has been described in the context of targeted studies of specific planets like Mars \cite{checinska2019} or Venus \cite{westall2023}, or in terms of a descriptive approach detailing each mechanism that can affect habitability and the atmosphere \cite[e.g.][]{gillmann2022}. Here, we instead draw from the core science questions that remain unsolved, summarise the current state of our knowledge and discuss how future work can build upon this foundation. 
We have identified four broad questions that are the pillars of ongoing investigations into the habitability of terrestrial planets in the Solar System and beyond:

\begin{itemize}
\item How do planets become habitable and to what degree is it influenced by their accretion history, bulk composition and initial conditions?

\item How do the interior dynamics of a planet and its mantle convection regime affect habitability?

\item What can cause habitability to end or push a planet out of the regulating cycles that maintain habitable surface conditions?

\end{itemize}

\section{The onset of habitability}
\label{sec:onset}

\subsection{Volatile Budgets}
\label{subsec:volatile_budget}

Both planetesimal- and pebble accretion imply a considerable degree of stochasticity as to the volatile element budgets of growing terrestrial planets \cite{albarede2009,sossi2022stochastic}. Planets accumulate material from a wide range of heliocentric feeding zones \cite{wetherill1994}, regardless of whether condensed material was initially confined to an annulus \cite{jacobsonmorbidelli2014} or ring \cite{izidoro_etal2022,morbidelli_etal2022} or whether it resulted from inward-drifting pebbles \cite{johansenlambrechts2017}. 
This stochasticity leads to a gradual decline in elemental abundance with increasing volatility (decreasing condensation temperature\footnote{The temperature at which half of the mass of a given element condenses from a gas of solar nebular composition, at a nominal pressure of 10$^{-4}$ bar.}) in the terrestrial planets 
\cite{sossi2022stochastic}. Elements that were sequestered into the core during its formation (siderophile elements) show depletions in the mantle with respect to the `volatility trend' defined by silicate-loving (lithophile) elements \cite[e.g.,][]{fegley2020volatile}. Therefore, although the total budgets of siderophile elements in the terrestrial planets are higher than those observed in their mantles, they are inaccessible to geochemical scrutiny, and do not play a direct role in determining habitability conditions at the surface of the planet \cite[although they may indirectly contribute, such as in the generation of a geodynamo,][]{nimmo2004influence}. \\

As such, the bulk planetary abundances of key atmosphere-forming moderately volatile elements, of which S and P are part, can be estimated from those of lithophile elements of similar volatility elements. 
Conversely, highly volatile elements, notably C, H and N, because of their low abundances in planetary mantles with respect to the nebular gas or chondritic meteorites \cite[e.g.,][]{mccubbinbarnes2019}, are likely to have been delivered irregularly and near the tail end of accretion \cite[e.g.,][]{albarede2009,marty2012}. As a result, their provenance is likely distinct from that of major, planet-forming elements \cite[see][]{dauphas2017, broadley_etal2022}, such that predicting their abundances in the terrestrial planets is fraught with uncertainty.\\

Through analysis of peridotitic rocks that comprise the Earth's mantle, together with inversion of the compositions of products of mantle melting in the ocean basins
, the concentrations of volatile elements in the convecting mantle of the Earth can be determined \cite[cf.][]{marty2012,palmeoneill2014,hirschmann2018}. The mass of oxygen is overwhelmingly contained in the Earth's mantle, bound in silicates and oxides, and can be computed by stoichiometry to better than 2 \% relative. The moderately volatile elements, P and S, also reside chiefly in the mantle, and have abundances that can be determined to a precision of $\sim$20 \% relative \cite{wangbecker2013,palmeoneill2014}. \\

, 
H, C and N are presumed to exist as a particular species (H$_2$O, CO$_2$, N$_2$) during mantle melting, and their abundances in basalts are normalised to those of a refractory lithopile element 
(H$_2$O/Ce CO$_2$/Ba or CO$_2$/Nb), or failing that, another volatile element (Ar for N$_2$) that behaves in a similar manner 
\cite[e.g.,][]{hirschmann2018}, with uncertainties of 
50 \% relative. Furthermore, significant fractions of the budgets of H, C and N are present on the surface \cite[atmosphere + oceans + crust;][]{rubey1951} of the Earth, known to within 20 \% relative. 
\cite[see][and references therein]{hirschmann2018}. Together, the abundances of H, C and N in the bulk silicate Earth (BSE) can be computed within a factor $\sim$2. With these caveats in mind, the depletion factors of the six key elements, CHONPS, in the BSE relative to the solar- and CI-chondritic compositions are found in Table \ref{tab:CHONPS_abundances}.\\

\begin{table}[!ht]
    \small{}
    \centering
    \caption{The abundances of the life-essential volatile elements, C, H, O, N, P and S, in the bulk silicate Earth (BSE), CI chondrites and in the Sun. All concentrations expressed in weight fraction. Uncertainties for C, H and N in the BSE are of the order of a factor $\sim$2, while P and S are $\pm$20 \%, with O $\pm$2 \%}
    \begin{tabular}{llllllll}
    \hline
        \textbf{} & \textbf{C} & \textbf{H} & \textbf{O} & \textbf{N} & \textbf{P} & \textbf{S } & \textbf{} \\ \hline
        BSE & 1.4$\times$10$^{-4}$ & 1.2$\times$10$^{-4}$ & 0.4433 & 2.8$\times$10$^{-6}$ & 8.7$\times$10$^{-5}$ & 2.0$\times$10$^{-4}$ \\ 
        CI chondrite & 0.0348 & 0.0197 & 0.459 & 2.95$\times$10$^{-3}$ & 9.85$\times$10$^{-4}$ & 0.0535 \\ 
        Sun & 2.17$\times$10$^{-3}$ & 0.738 & 6.34$\times$10$^{-3}$ & 7.48$\times$10$^{-4}$ & 6.44$\times$10$^{-6}$ & 3.42$\times$10$^{-4}$  \\ 
        Depletion (/CI, Al) & 0.0014 & 0.0022 & 0.341 & 3.3$\times$10$^{-4}$ & 0.031 & 0.0013 \\ 
        Depletion (/Sun, Al) & 1.56$\times$10$^{-4}$ & 3.93$\times$10$^{-7}$ & 0.169 & 9.04$\times$10$^{-6}$ & 0.033 & 0.0014 \\ \hline
    \end{tabular}
    \label{tab:CHONPS_abundances}
\end{table}

A striking observation is that no more than $\sim$0.4 ppm (by weight, ppmw) of solar nebula matter remains in the Earth, based on the Al-normalised abundance of H (Table \ref{tab:CHONPS_abundances}). Of course, this estimate ignores any amount of H potentially dissolved in the core.  
Assuming a partition coefficient between core and mantle, D$^H_{core/mantle}$ = 10 \cite{li_etal2020Hcore}, then the 120 ppmw H in the mantle would give 0.12 wt. \% in the core, consistent with geophysical constraints \cite{hirose_etal2021}. 
Even if D$^H_{core/mantle}$ = 100 \cite[as permitted by the computations of][]{li_etal2020Hcore} and 1.2 wt. \% H were to be dissolved in the core, the depletion factor of H relative to Al and the solar nebula (Table \ref{tab:CHONPS_abundances}) would be $\sim$10$^{-5}$, lower than or equivalent to, respectively, the solar-, Al-normalised abundances of C or N, even without accounting for the siderophile behaviour of the latter pair \cite[e.g.,][]{speelmanns_etal2019,grewal2022,blanchard_etal2022metal}. 
\\

Therefore, it seems that, compared to the Sun, the Earth (or its building blocks) underwent preferential depletion of H relative to similarly volatile elements, C and N, with the noble gases having been depleted by an even greater factor than H \cite{aston1924,fegleyschaefer2014}. Indeed, \cite{aston1924} recognised that \textit{``the earth has only one millionth part of its proper quota of inert gases"}, with modern estimates \cite{marty2012,halliday2013} ranging from 1 part in 10 billion (Ne) to one one-millionth (Xe). Hence, the Earth has retained but a minuscule fraction of a solar component (i.e., from the nebular gas). \\

The origin of the small fraction of volatiles has been reconciled with either \textit{i)} dissolution and subsequent loss of a component of the nebular gas \cite[e.g.,][]{olson2019nebular} or \textit{ii)} predominantly dry accretion of the Earth with a small component of volatile-rich, chondritic and/or cometary material \cite[e.g.,][]{albarede2009}. Isotope ratios among volatile elements offer a complementary way to distinguish between these plausible sources \cite[see][for a review]{peron2018origin}.  
The isotopic compositions of noble gases (except He and Ne), H, C and N in the BSE overlap with those of chondrites \cite{alexander_etal2012,piani_etal2020,broadley_etal2022}, which, themselves, are distinct from that of the solar nebula (as inferred from the Sun and the atmospheres of the gas giants). \\

Models of dissolution of the solar nebula can account for the observed abundance of H relative to the noble gases \cite{sharp2022multi,young2023earth}. \cite{sharp2022multi} highlight the plurality of sources that could have been accreted to the Earth to satisfy both elemental and isotopic constraints, with N and heavy noble gas abundances being consistent with chondritic and cometary sources, respectively, whereas H and the light noble gases permit a greater fraction of nebular material. Yet, such a process would have resulted in an overabundance relative to the BSE (Table \ref{tab:CHONPS_abundances}) and an H isotope composition far too light (D/H of $\sim$30 $\times$ 10$^{-6}$), necessitating isotopic fractionation associated with its subsequent loss via atmospheric escape. These models have, so far, not considered carbon in their treatment, making it unclear whether dissolution of the nebular gas is capable of producing a match to the HCN abundances of the BSE. \\

 \begin{figure}[!ht]
     \centering
     \includegraphics[width=0.75\textwidth]{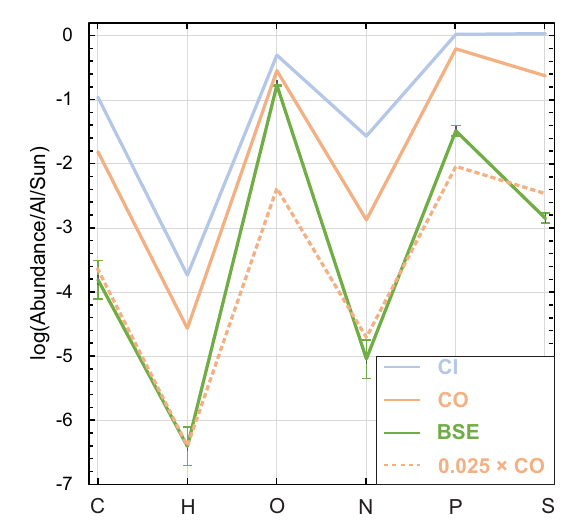}
     \caption{The abundances of the life-essential volatile elements, CHONPS, relative to the solar abundances \cite{lodders2010} normalised by Al in: CI chondrites \cite[teal line,][]{palmeoneill2014}, CO chondrites \cite[orange line,][]{alexander_etal2012} and the bulk silicate Earth \cite[BSE; green line,][]{marty2012,palmeoneill2014,hirschmann2018}. The dashed orange line represents 2.5 \% of the CO abundances.}
     \label{fig:major_volatiles_BSE}
 \end{figure}

On the other hand, to first order, the relative abundances of volatile elements bear a strong resemblance to those in chondrites \cite[][Fig. \ref{fig:major_volatiles_BSE}]{broadley_etal2022}, suggesting that the present-day budgets of these elements were brought by such materials 
\cite{piani_etal2020,marty2022}, although Ne isotopes permit a small fraction 
of its budget in the BSE to be solar in origin \cite{yokochimarty2004,williamsmukhopadhyay2019,sharp2022multi}. However, it should be kept in mind that this is relative to the 10$^{-7}$ that remains of the Ne budget of the solar nebula.  
\\

The precise nature of the chondritic material that could have delivered these volatiles, and, what fraction of their total budgets were contributed by such material, remains under scrutiny. 
Assuming that the D/H and N isotopic compositions of the BSE reflect those of their sources, the Earth's H and N (and, by extension, C) budgets were thought to have been carried by CI- or CM-type carbonaceous chondrites \cite[e.g.,][]{alexander_etal2012,furimarty2015}. This interpretation has been called into question on the grounds that the Earth is known to share an isotopic heritage with enstatite chondrites (EC) for the major rock-forming elements \cite[e.g.,][]{dauphas2017}. Consequently, \cite{piani_etal2020} postulated that EC-like material could have delivered much of the H budget of the Earth. 
A major caveat is that D/H and N isotope ratios could be fractionated in a mass-dependent manner \cite{grewal2022,sharp2022multi}, and thus not only record the provenance of these elements, but also mass transfer processes (such as atmospheric escape) that could have modified their terrestrial inventories thereafter. 
\\

Figure \ref{fig:major_volatiles_BSE} indicates that the addition of 2.5 \% CO-chondritic material by mass to an otherwise volatile-free Earth would reproduce the present-day H, C and N budgets of the BSE, within uncertainty. 
Equally, 0.7 \% of CI-like material, consistent with the mass of the late veneer \cite[cf.][]{albarede2009}, would satisfy the H budget of the Earth, but would deliver 6 times too much N. 
However, the observation that H, C and N are depleted relative to the noble gases \cite[normalised to CI chondrites;][]{halliday2013} suggests that a straightforward chondritic contribution is too simplistic, and that these elements preserve some vestige of the main accretion phase. 
Therefore, in the event that chondritic material delivered all of the H, C and N, this would have represented, at most, $\sim$10--25 \% and 2.5\% of the P and O budgets of the BSE, respectively. Hence, P and O must have delivered during the main accretion phase of the Earth and hence from predominantly non-carbonaceous material \cite[e.g.,][]{dauphas2017}. The addition of 0.7 \% CI or 2.5 \% CO-like material would exceed the S budget of the BSE by 1.5--2.5 $\times$. However, S, together with Se and Te, was likely depleted by the extraction of a small amount of sulfide-rich melt from the mantle \textit{after} the cessation of core formation, but \textit{before} the accretion of the late veneer \cite{oneill1991Earth,wangbecker2013}. \\

In sum, the bulk volatile inventory of terrestrial planets, taking the Earth as an example, is low ($\leq$ 0.05 \% by mass). 
This observation, together with the chondritic relative abundances and isotope ratios of major volatiles, indicates  chondritic material (although of uncertain origin) as the major source of H, C and N in the BSE. Their isotopic and elemental compositions are more difficult to reconcile with capture and dissolution of the nebular gas, without appealing to additional processes (i.e., atmospheric escape) and would require additional investigation. %
Instead, it points towards accretion via a mixture of largely volatile-depleted bodies with smaller amounts 
carbonaceous chondrites \cite[though enstatite chondrites remain feasible][]{piani_etal2020}, representing $\geq$30 \% of the present-day BSE budgets of H, C and N \cite[see also][]{steller_etal2022}. Budgets of the less volatile P ($\leq$25 \% CC-material) and O ($\leq$2 \% CC-material) were brought predominantly by sources other than carbonaceous chondrites, whereas S, 
likely reflects the composition of the late veneer.

\subsection{Volatile Distribution}
\label{subsec:volatile_distribution}

The constitution of an atmosphere, during a magma ocean epoch, is dictated by the abundances of elements available in planet-building material (see section \ref{subsec:volatile_budget}), the intensive variables, namely temperature and pressure that determine the speciation of gaseous molecules, and the solubility relations these gaseous species have with their dissolved counterparts in silicate liquids. Because the masses of atmospheres around terrestrial planets are small relative to their interiors, the assumption is frequently made that the interior imposes the state of the atmosphere. As such, the partial pressure (\textit{p}) of given species, \textit{i}, in the atmosphere is:

\begin{equation}
p_i = {x_i} \frac{g M^a}{4 \pi r^2} ,
\label{eq:sol1}
\end{equation}

where \textit{g} = the acceleration due to gravity, $M^a$ is the mass of the atmosphere, \textit{$x_i$} is the \textit{molar} fraction of element i and \textit{r} is the planetary radius. To determine the value of $M^a_i$ requires knowledge of the distribution of a given gas species in the magma ocean using a solubility law. A solubility law relates the fugacity of species \textit{i}, \textit{f$_i$}, (or partial pressure, given an ideal gas expected to typify most secondary atmospheres around terrestrial planets) to the mole fraction, \textit{X}, of the relevant species dissolved in a condensed phase, here, a silicate liquid;

\begin{equation}
X_i = \alpha f_i^{\beta} ,
\label{eq:sol}
\end{equation}

where $\alpha$ is a Henrian constant and $\beta$ is the stoichiometric coefficient. The relationship between the fraction of mass of species $i$ in the atmosphere (M$^a_i$) relative to that in the mantle (M$^m_i$), assuming that the sum of the two is equivalent to the total mass, M$_i^T$ = M$^a_i$ + M$^m_i$, and that \textit{f$_i$} = \textit{p$_i$} with $\beta$ = 1, is given by:

\begin{equation}
\frac{M^a_i}{M^m_i} = \frac{4 \pi r^2}{g M_T \alpha},
\label{eq:solubility_dependence}
\end{equation}

where $M_T$ is the total mass of the planet. Taking a generic chemical reaction and setting $\alpha$ = 1000 ppm/bar (in SI units, 10$^{-8}$ Pa$^{-1}$) and $\beta$ = 1, the distribution of a volatile species, comprising 1000 ppm of the planet, between the mantle and atmosphere is determined as a function of planetary mass, $M_T$, where \textit{g} is given at known \textit{r} by solving the Adams-Williamson equation of state \cite[see][]{Valencia:2007jv,sossi_etal2023}. \\

\begin{figure}[!ht]
     \centering
     \includegraphics[width=0.75\textwidth]{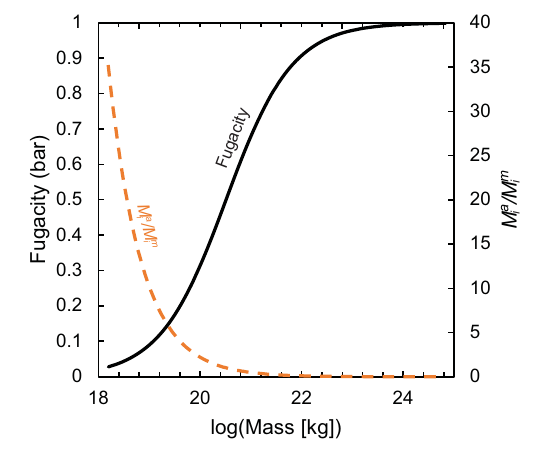}
     \caption{The distribution of a fictive element, \textit{i}, with $\alpha$ = 1000 ppm/bar and $\beta$ = 1 (cf. eq. \ref{eq:sol}) between the molten interior (\textit{m}) and atmosphere (\textit{a}), expressed in terms of its mass \textit{M$_i$} and fugacity (in bar) as a function of the log$_{10}$ of the planetary mass in kg. The planet is assumed to be Earth-like, in which an silicate mantle comprises 70 \% of its mass, with the remaining 30 \% being made up of an Fe-rich core.}
     \label{fig:solubility_distribution}
 \end{figure}

Figure \ref{fig:solubility_distribution} shows that Eq. \ref{eq:solubility_dependence} predicts that the quantity M$^a_i$/M$^m_i$ decreases as planetary mass (proportional to \textit{g}) increases. That is, as shown by \cite{sossi_etal2023}, small, rocky bodies will have a greater fraction of a given volatile element present in the atmosphere relative to the molten interior as compared to larger planets. 
Despite the decreasing \textit{relative} fraction of a given volatile element in the atmosphere, its partial pressure (or fugacity) in the atmosphere increases asymptotically to a maximum value given by Eq. \ref{eq:sol}; 
$f_i$ = $X_i$/$\alpha$. Consequently, mass is a key property that dictates whether the planet is capable of harbouring an atmosphere. \\

Although some species, notably N$_2$ \cite[at high \textit{f}O$_2$][]{libourel_etal2003nitrogen}, CO$_2$ \cite{dixon_etal1995}, H$_2$O \cite[at high \textit{f}H$_2$O;][]{stolper1982water} and H$_2$ \cite[at high \textit{f}H$_2$;][]{hirschmann_etal2012} dissolve in silicate melts with $\beta$ $\sim$ 1, others have more complicated dissolution stoichiometries. At low $f$H$_2$O ($\leq$ 1000 bar), water (and possibly H$_2$ at low \textit{f}H$_2$), dissolves into silicate melts with $\beta$ = 0.5 \cite{newcombe_etal2017,sossi_etal2023}. At low \textit{f}O$_2$ nitrogen dissolves in silicate melts as nitride, N$^{3-}$, a species that is more soluble than the diatomic molecule \cite{libourel_etal2003nitrogen,bernadou_etal2021}. The solubility of sulfur has long been known to be redox-sensitive \cite{finchamrichardson1954,oneillmavrogenes2002sulfide,boulliungwood2022so2}, existing in silicate liquids as S$^{2-}$ at low \textit{f}O$_2$ and SO$_4 ^{2-}$ at high \textit{f}O$_2$. \\   

In this context, the oxygen fugacity (\textit{f}O$_2$) holds special significance, as, unlike the other CHONPS elements, O is invariably in excess in the mantle of the planet (and in excess relative to the other major volatile elements; Table \ref{tab:CHONPS_abundances}) with respect to the atmosphere. Consequently, at equilibrium, the \textit{f}O$_2$ of the atmosphere is equivalent to that in the mantle, given by:

\begin{equation}
FeO(l) + \frac{1}{4}O_2(g) = FeO_{1.5}(l) ,
\label{eq:Fe-fO2}
\end{equation}

\cite{sossi2020redox} exploited this property to determine the speciation and mass of an atmosphere in equilibrium with a magma ocean of peridotitic composition; assuming Fe$^{3+}$O$_{1.5}$/(Fe$^{3+}$O$_{1.5}$ + Fe$^{2+}$O) = 0.037 (that of the present-day Earth's mantle), would set the \textit{f}O$_2$ of the atmosphere in equilibrium with the magma to $\sim$IW\footnote{IW = the $f$O$_2$ set by the iron-wüstite buffer; FeO = Fe + 1/2O$_2$.} \cite[see also][]{hirschmann2022magma}. At 2173 K, the temperature at which peridotite is fully molten at 1 bar pressure, this corresponds to $\sim$10$^{-6}$ bar of O$_2$. Its importance lies in fixing the relative fugacities of atmospheric species that can be related by \textit{f}O$_2$ alone, such as; 

\begin{subequations}
\begin{equation}
H_2(g) + \frac{1}{2}O_2(g) = H_2O(g),
\label{eq:H2-H2O}
\end{equation}
\begin{equation}
CO(g) + \frac{1}{2}O_2(g) = CO_2(g),
\label{eq:CO-CO2}
\end{equation}
\end{subequations}

The CO/CO$_2$ and H$_2$/H$_2$O ratios of the atmosphere, as well as their total pressures, are key in dictating the physical and chemical stability of atmospheres around rocky planets \cite[e.g.,][]{lichtenberg2021vertically}. The H$_2$/H$_2$O ratio, in particular, influences the fugacities of CH$_4$ and NH$_3$, both of which are instrumental in the production of amino-acids by spark discharge in the presence of liquid H$_2$O, as investigated in the Miller-Urey experiment \cite{miller1953}. Their homogeneous gas phase reactions can be written;

\begin{subequations}
\begin{equation}
CO(g) + 3H_2(g) = CH_4(g) + H_2O(g),
\label{eq:CH4}
\end{equation}
\begin{equation}
N_2(g) + 3H_2(g) = 2NH_3(g),
\label{eq:NH3}
\end{equation}
\end{subequations}

\begin{figure}[!ht]
     \centering
     \includegraphics[width=1\textwidth]{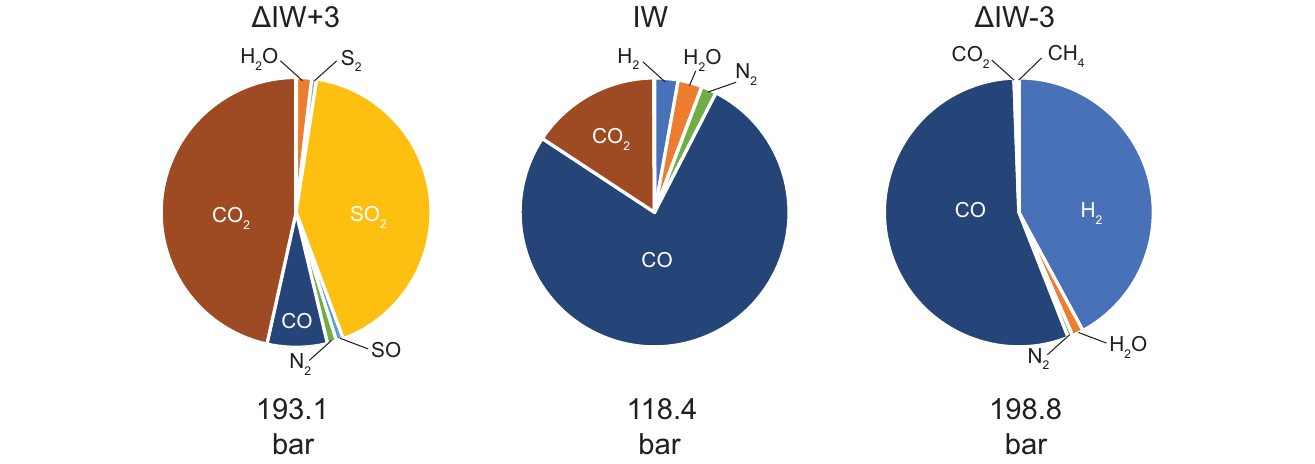}
     \caption{Equilibrium speciation of atmospheres at 2173 K formed around an Earth-sized planet in equilibrium with a magma ocean of mass equal to that in Earth's mantle (4.2 $\times$ 10$^{24}$ kg) in the system CHONPS with abundances for the bulk silicate Earth (BSE) as per Table \ref{tab:CHONPS_abundances}. The atmospheres are produced given solubility laws for CO \cite{yoshioka_etal2019}, CO$_2$ \cite{dixon_etal1995}, H$_2$O \cite{sossi_etal2023}, H$_2$ \cite{hirschmann_etal2012}, N$_2$ \cite{libourel_etal2003nitrogen}, SO$_2$ \cite{boulliungwood2022so2} and S$_2$ \cite{boulliungwood2023sulfur} and their equilibrium gas speciation solved at three oxygen fugacities relative to the iron-wüstite (IW) buffer, +3, 0 and -3. Total pressure in the atmosphere is shown below the respective pie chart.}
     \label{fig:Atmospheres-2173K}
 \end{figure}

The prevalence of such gases depends on \textit{i)} the absolute abundances of H, C and N in the atmosphere and \textit{ii)} the atmospheric H$_2$/H$_2$O ratio, which is, in turn, governed by the temperature, pressure and \textit{f}O$_2$. In order to investigate the effect of \textit{f}O$_2$ on the chemistry of atmospheres produced on the early Earth in equilibrium with a magma ocean, the BSE abundances of the CHONPS elements (Table \ref{tab:CHONPS_abundances}) are used together with the relative fugacities of stable gas species given by the equilibrium constants (Chase, 1998) of their homogeneous gas phase reactions (e.g., eqs. \ref{eq:H2-H2O}, \ref{eq:CO-CO2}, \ref{eq:CH4}, \ref{eq:NH3}) and their solubilities in silicate liquids \cite[see][and references therein]{bower_etal2022}. The partial pressures (fugacities) of the major atmosphere-forming species at 2173 K at three different oxygen fugacities \cite[expressed relative to the iron-wüstite buffer;][]{oneillpownceby1993} in equilibrium with an Earth's mantle-sized magma ocean are shown in Fig. \ref{fig:Atmospheres-2173K}. \\

Under reducing conditions ($\Delta$IW-3), an early Earth atmosphere would have had subequal amounts of CO and H$_2$ \cite[cf.][Fig. \ref{fig:Atmospheres-2173K}]{sossi2020redox}, while CH$_4$ and NH$_3$ are present at 0.35 bar and 0.013 bar, respectively. By contrast, S is essentially absent, with the most abundant S-bearing species, H$_2$S = 2$\times$10$^{-5}$ bar. The total pressure at the IW buffer defines a minimum, $\sim$120 bar, as, at lower \textit{f}O$_2$, CO is joined by H$_2$ and at high \textit{f}O$_2$, CO is replaced by CO$_2$ and significant quantities of SO$_2$ ($\sim$80 bar, but also $\sim$1 bar each of SO and S$_2$) are present in the atmosphere, a result emphasised by \cite{gaillard_etal2022}. This reflects the low solubility of SO$_4^{2-}$ in silicate melt at high \textit{f}O$_2$ \cite[cf.][]{boulliungwood2023sulfur}. 
Water is not a major component in any atmospheric type as its high solubility ensures it remains dissolved in the silicate liquid \cite[see also][]{bower_etal2022,sossi_etal2023}, with the implication being that the $\sim$270 bar equivalent mass of H$_2$O present in the oceans today is a product of the long-term evolution of the Earth (see section \ref{sec:conv}), starting with the solidification of the magma ocean \cite{bower_etal2022,salvador2023,maurice2023}. That is, owing to the lower solubility of H$_2$O in solid phases relative to silicate liquid, crystallisation tends to expel water from the magma ocean into the atmosphere. However, the degree to which H$_2$O degasses depends upon the mode of crystallisation, where convective lock-up of the magma ocean when the melt fraction is $\sim$0.3 prevents further outgassing of water, whereas efficient crystal settling leads to a surface magma ocean, allows water to escape \cite{bower_etal2022}. The sum of partial pressures of phosphorus-bearing species are low, ranging from 2$\times$10$^{-3}$ bar at $\Delta$IW-3 (mainly P, PO$_2$, PO and P$_2$) to 3$\times$10$^{-6}$ bar at $\Delta$IW+3 (chiefly PO$_2$). Their low contribution results from the non-ideality of P$_2$O$_5$ in silicate melts \cite[$\gamma$P$_2$O$_5$ $\sim$10$^{-10}$;][]{suito1981phosphorus,fegley2023chemical} and the P abundance in the BSE (87 ppmw; Table \ref{tab:CHONPS_abundances}). \\

The relative abundances of gaseous species (Fig. \ref{fig:Atmospheres-2173K}), change markedly as the atmospheres cool and components condense and are extracted from the gas phase \cite{sossi2020redox,liggins_etal2022}. Notably, CH$_4$ and NH$_3$ become increasingly stable down-temperature at the expense of H$_2$ and CO. However, the magma ocean stage is likely to supply the initial budgets of elements to the surface that may then become available to prebiotic synthesis pathways. 
Greater fractions of S and N are present in more oxidising atmospheres, whereas P- and H-bearing species are stabilised in the gas phase at reducing conditions, while C remains volatile throughout the \textit{f}O$_2$ range explored. As such, essential prebiotic molecules, such as HCN, can only form when the partial pressures of H$_2$, CO and N$_2$ are high enough, with its presence being favoured at reducing conditions, together with CH$_4$ and NH$_3$. This realisation led to the notion that transient impact events on the early Earth could have locally produced reducing atmospheres \cite{Zahnle_2020,Itcovitz_etal2022,thompson_etal2022,wogan2023origin}. Nevertheless, whether these gases can contribute to prebiotic reactions hinges upon the photochemical stability of these species as atmospheres around rocky planets cool over time.

\section{The role of the convection regime}
\label{sec:conv}

\subsection{Convection regimes On Earth and other planets}
Following accretion, rocky planets cool down and crystallize from a magma ocean, leading to the onset of solid-state mantle convection. Different interior dynamics give rise to distinct surface expressions, including habitability. Mantle convection and its surface expressions have a profound impact in the thermal and compositional evolution of a planet, as mantle dynamics link various layers of the planet and drive evolution. In particular, it effectively links the deep interior (down the the planetary core) with the surface/atmosphere through heat transfer and the effect of the magnetic field (see section \ref{sec:mag}). Other main expressions of this link include volcanism, and the resulting outgassing of volatile species into the atmosphere, that in turn affects atmospheric mass and surface conditions through the greenhouse effect. In conjunction with other processes (photochemistry, escape, for example), they contribute to the evolution of planetary surface conditions. Together, they effectively produce a coupled system with complex feedback processes, which we are only starting to grasp \cite[i.e.][]{schubert1989coupled,lenardic2008climate,driscoll2013,gillmann2014atmosphere,gillmann2022,krissansen2021,ortenzi2020,foley2016}. Here, we focus on the internal component of the system.

\begin{figure}[!ht]
     \centering
     \includegraphics[width=1\textwidth]{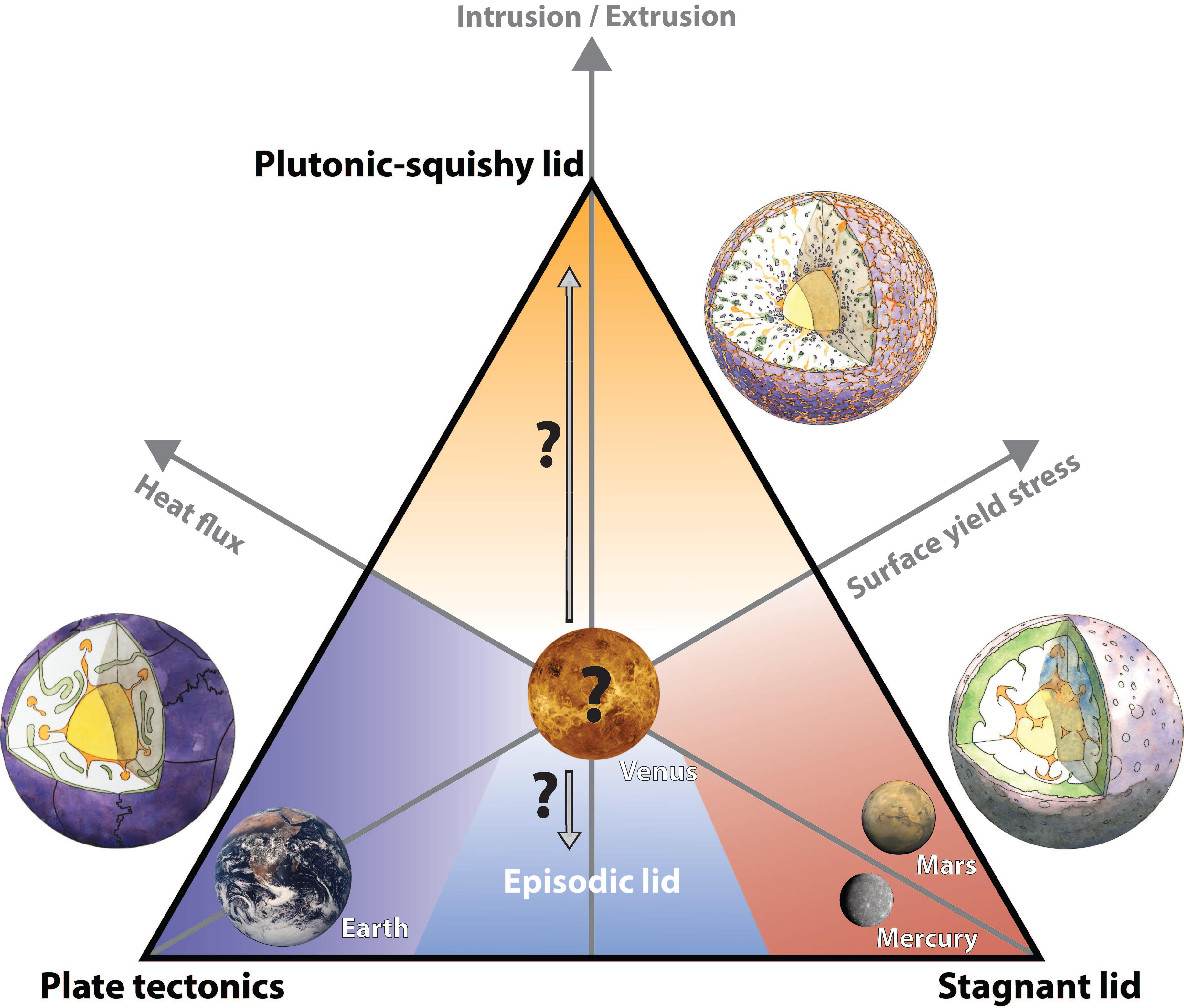}
     \caption{Schematic depiction of the primary convection regimes observed or theorized on terrestrial planets in the Solar System. The three end-members, in this representation, are Plate Tectonics (PT), Stagnant Lid (SL) and Plutonic-Squishy Lid (PSL). Episodic Lid (EL) is an intermediate regime between PT and SL. While the current regimes of Earth, Mercury and Mars are well defined, Venus' is still uncertain and could range from EL to PSL, or may have transitioned recently (a few 100s Myr) into SL.  Mercury is in a later stage of SL compared to Mars because it has cooled down significantly faster. The state of the convection regimes of the planets should not be considered static. Instead, it evolves with time, and the figure can be interpreted as a snapshot of their present-day state. The axis are not fully independent and should be thought of as possible observables. The convection regime figures are adapted from \cite{Lourenco2023}.}
     \label{fig:convectionR}
 \end{figure}

Various convective or tectonic regimes have been proposed to exist for different interior states (see Fig. \ref{fig:convectionR}, aligning with distinctive observed features. Notably, the Earth presently exhibits plate tectonics \cite{wegener1915entstehung,mck67,morgan68,pichon68}: namely, its outermost rigid shell, the lithosphere, is fragmented into several sizable tectonic plates, gradually shifting atop a mechanically weaker layer called the asthenosphere (part of the upper mantle). Along convergent boundaries, at subduction zones, cold and dense oceanic plates sink into the mantle. This subducted material is balanced by the generation of new oceanic crust along divergent margins and the growth of the lithosphere due to heat diffusion. These dynamics are intrinsically related with mantle convection, i.e. the gradual, creeping motion of the Earth's solid mantle. 

Advances in numerical modeling have significantly enhanced our understanding of plate tectonics, while also revealing other tectonic regimes that can explain observations from other Solar System bodies. These alternative regimes include the stagnant-lid (as observed on Mars, Mercury or the Moon), where a planet is covered by a single plate \cite{Nataf:1982gc,CHRISTENSEN:1984bq,Solomatov:1995ki}, the episodic-lid regime, characterized by a predominantly motionless lithosphere occasionally undergoing overturning into the mantle \cite{Turcotte:1993ko, Moresi:1998hr, Stein:2004vp, Armann:2012ce,gillmann2014atmosphere,Lourenco:2016kl}, the ridge-only regime, where a large ridge emerges, slightly compressing the surrounding lithosphere without forming subduction zones \cite{Tackley:2000en, Rozel:2015jw}, and the plutonic-squishy lid regime, distinguished by a thin lithosphere and several small, strong plates or blocks separated by warm and weak regions generated by plutonism \cite{rozel2017continental,lourencoPSL2020}. Venus is currently believed to have either an episodic or a plutonic-squishy lid, yet the exact nature of the tectonic regime on Venus remains mysterious \cite{gillmann2022,rolf2022dynamics}.

On Earth, substantial evidence supports the notion that plate tectonics played a pivotal -perhaps indispensable- role in the genesis and development of its atmosphere, oceans, continents, and life \cite[e.g.,][]{dehant2019geoscience, sobolev:2011nat, Stern:2016cc, zaffos2017,stern2023}. Increasing evidence indicates that plate tectonics, as a global tectono-magmatic system, has two significant impacts: (1) it governs the cycling of essential nutrients between various reservoirs such as the geosphere, oceans, and atmosphere through diverse tectonic, magmatic, and surface processes \cite{Zerkle2018}, and (2) the gradual 
movement of tectonic plates, along with associated topographic alterations and consistent moderate magmatic activity, shaped and fostered biological evolution \cite[e.g.,][]{leprieur2016, pellissier2017, Zerkle2018}.

It is worth pointing out that there are indications that the inception of life might have occurred on an Earth 
devoid of plate tectonics. This proposition stems from zircon palaeointensity data, which indicates unvarying latitudes for the samples dating back from about 3.9 billion years ago (Ga) to about 3.3 Ga \cite{tarduno2023}, implying that no large-scale surface motion occurred at the time on Earth \cite[contrast for example with][and references therein]{lammer2018}. The earliest indications of life on Earth are older than 3.7 Ga \cite[e.g.,][]{rosing99}.

Despite its significance on Earth, plate tectonics has not been observed on any other planet. Over the last few years, there has been a notable increase in the discovery of rocky exoplanets, particularly the ones known as super-Earths, which are large terrestrial planets with masses up to ten times that of Earth, roughly up to twice its diameter. 
The majority of studies indicate that these planets should experience a mobile-lid regime or plate tectonics \cite{Valencia:2007jv,Valencia:2009ij,vanHeck:2011df,Tackley:2013cs}.
Nevertheless, some suggest the contrary \cite{ONeill:2007do,Stein:2011hy}. Therefore, it is also worth contemplating whether life can emerge and flourish on planets with different tectonic regimes.

\subsection{The interior of terrestrial planets and outgassing}
The study of how the interiors of terrestrial planets relate to their outgassing histories is still in its early stages.  
Surface conditions and atmospheric composition are controlled by outgassing (together with volatile sinks like atmospheric escape and weathering), while volcanism and tectonics share a profound connection. The case of Mars illustrates that stagnant lid planets could outgas sufficient volatiles to affect the evolution of their surface conditions \cite{gillmann2009present,gillmann2011volatiles,grott2011volcanic,leblanc2012mars,lammer2013outgassing,ramirez2014warming}.
However, it is often suggested that, on average, volcanism on a planet with plate tectonics would lead to more pronounced outgassing than on a stagnant lid (albeit not necessarily more melt production, see \cite{Armann:2012ce}). 
On the other hand, an episodic lid regime \cite{gillmann2014atmosphere} could produce a step-like evolution of surface conditions, with changes occurring in short, sudden bursts, which may be detrimental to life. At an extreme level, some volcanic events, such as Large Igneous Provinces have been suggested to alter surface conditions so strongly that they could cause the end of planetary habitability \cite[see][for Venus]{way2022large}.

The composition and release of gases during partial melting and volcanic activity depend on the mantle's composition and redox state (section \ref{sec:onset}). However, uncertainties exist regarding these factors, even for Earth \cite{berry2008,canil97,canil2002,kellergeochemistry2012,gaillard2015,Aulbach2016,sossi2020redox}. The redox state influences the speciation of volatile elements in magmas, and, given that these species have different solubilities, the degree to which they are released from magmas (cf. \ref{sec:onset}, eqs. \ref{eq:H2-H2O}, \ref{eq:H2-H2O}). 

The redox state of the mantle (\textit{f}O$_2$) is likely inherited from the composition of planetary building blocks, and the conditions under which core-mantle equilibrium took place \cite[e.g.][]{armstrong_etal2019,deng2020magma}. However, these initial conditions  may be affected by the subsequent tectonic regime experienced by a planet. Notably, it has been proposed that subduction, through the burial of oxidised components such as H$_2$O, CO$_2$ and sulfates, has lead to an increase in the oxidation state of Earth's mantle over time, although this remains a topic of debate \cite[e.g.,][]{Aulbach2016,stolper2019,nicklas2018redox}, while other processes, such as disproportionation of Fe$^{2+}$ into Fe$^{0}$ and Fe$^{3+}$ are also plausible \cite{frostmccammon2008,armstrong_etal2019}.

The pressure at which degassing occurs also plays a significant role in determining the composition and mass of the released species \cite{gaillard2014}. In the case of Venus, with a surface pressure $\sim$90 times greater than Earth's, volcanoes are expected to emit carbon-dominated gases \cite{gillmann2022}, whereas on Earth, volcanoes release more water-rich gases. For large exoplanets with extremely high atmospheric pressures, fewer volatiles would be released 
\cite{gaillard2014,gaillard21}.

The size of a planet also also appears as a major factor influencing volatile history. For instance, escape processes depend in a large part on the gravitational pull of the planet, with small planets experiencing stronger loss over time than larger ones \cite[for example,]{lammer2018}. One must also consider that different species escape at different rates. Lighter species tend to escape more readily; for instance Hydrogen. Heavier species tend to be depleted to a smaller extent. However, energy input from the Sun, atmosphere structure and composition, and photochemical reactions involving each species are critical to quantifying accurately the escape rates of each atmospheric component \cite{gronoff2022,wordsworth2022}. This affects volatile retention and the whole volatile inventory, from accretion to the present-day. The specifics of the escape processes vary from planet to planet, not only with size but also atmosphere composition and structure, as well as the presence or absence of a magnetic field (see section \ref{sec:mag}). This goes well beyond the scope of this work, and the interested reader is welcome to peruse further review papers, for instance: \cite{gronoff2022, wordsworth2022}, \cite{gillmann2022} (for a look at the interactions leading to the general volatile inventory) or \cite{salvador2023} (for the primitive evolution).

Focusing on internal processes, larger planets also generate more melt over longer periods. 
Because the heat of formation scales with volume and the heat loss scales with area, larger planets cool more gradually compared to small planets. 
The composition of the mantle and crust also affects the thermal evolution. For example, stars with C/O$>$0.65 \cite[for about 30\% of stars in the GALAH database,][]{spaargaren2023} are expected to be hosting carbon-rich rocky exoplanets \cite{bond2010,moriarty2014} with a graphite lid/crust or a diamond-silicate mantle. The presence of graphite/diamond would speed up planetary cooling by 10--100 times, leading to convection modes that are sluggish or stagnant with a low degree of partial melting \cite[][]{unterborn2014,hakim2019}. This difference arises from the 10--100 times higher thermal conductivities of graphite and diamond than that of common silicates. The final regime of planets that have lost most of their heat is an inactive stagnant lid (the Moon, Mercury), although not every stagnant-lid planet is geologically "dead".
The type of convection or tectonic regime they undergo is also crucial \cite{Lourenco:2018bp,lourencoPSL2020}. Recent studies have suggested that for stagnant-lid planets, melt production and volatile outgassing are more favorable for planets within the range of 2–4 Earth masses \cite{noack17,dorn18,ortenzi2020}.

\subsection{Observables}
Remaining unknowns regarding the interplay between degassing and tectonic regimes still impede efforts to reliably link specific atmospheres types (compositions, species, mass) to a given regime. 
Future work should aim at understanding the types of atmospheres that different tectonic regimes can build and their potential for detection.

Magmatic heat flow has been analyzed for various tectonic regimes \cite{lourencoPSL2020}. Heat flow is closely linked to erupted materials and, consequently, to outgassing. Provided sufficient heat sources, the highest magmatic heat flows are observed among active stagnant-lid planets with volcanism such as Io, which is the most volcanically active body in our Solar System \cite{Breuer2007}. Io's volcanism efficiently transports approximately 40 times the Earth's heat flux from its interior to its surface \cite{OReilly:1981es,Veeder:2004kh}. On the opposite end of the spectrum, plate tectonics and plutonic-squishy lid planets experience the lowest magmatic heat flows. Despite significant differences and uncertainties regarding intrusion to extrusion ratios, these regimes are expected to have relatively similar magmatic activity and outgassing histories compared to other tectonic regimes. Reliable measurements of the heat transport (through surface heat flow) are a major tool of planetary characterization of internal structure and dynamics \cite{smrekar2018venus,rolf2022dynamics,plesa2018thermal}, but remain delicate, even on Earth \cite{staal2022properties}.

\subsection{Surface and interior water inventory}

Assessing planetary habitability involves understanding the water inventory and its evolution over time. 
The planet's water retention depends on its evolution during the magma ocean stage, as some or most of the water is retained. On Earth, a significant portion of the delivered water was retained, becoming a fundamental cornerstone in its evolutionary history, which may not have been the case for Mars or Venus \cite[e.g.][]{gillmann2020,gillmann2022}. Collisions could also have affected the planetary climate \cite{segura2002environmental,turbet2020environmental,gillmann2016effect}.

Tectonic regimes also govern the availability of liquid water on the surface, as well as the weathering of silicate rocks and the burial of marine carbonate rocks into the mantle. These latter mechanisms contribute to climate cooling by removing CO$_{2}$ from the atmosphere and the planet's surface, respectively \cite[e.g.,][]{west2005, plank2019}. In turn, the recycling flux can deliver back volatiles to the mantle that can later be incorporated again in the magma. This mechanism could help prevent depletion of the mantle and ensure continuous outgassing over geological times.

The presence of water has far-reaching effects on mantle rheology, reducing its effective viscosity by several orders of magnitude \cite{katayama2008,karato2015}. It also influences melting processes by lowering the mantle solidus temperature \cite[e.g.,][]{green2010water}. Surface water plays a crucial role as well:

\begin{enumerate}
    \item Hydrated crust facilitates the lubrication of subduction slabs, enabling continuous one-sided oceanic subduction \cite{gerya2018,Crameri:2012ew}, a key factor and primary driver of plate tectonics \cite{forsyth1975}.
    \item On Earth, after glaciations, the delivery of sediments into oceanic trenches is enhanced, further lubricating subduction and accelerating plate tectonics \cite{Sobolev:2019ks}.
    \item As a condensable greenhouse gas, water significantly affects surface conditions, up to to runaway greenhouse (refer to section \ref{sec:cess}, \cite{goldblatt2012runaway}).
    \item The extent of Earth's surface covered by water significantly influences the planet's climate \cite{kodama2018,kodama2019,kodama2021,way2021,zhao2021,li2022,yang2020}, which can be linked to the thermo-convective evolution of Earth \cite{seales2020}.
\end{enumerate}

Therefore, changes in surface conditions affect the planet's interior. For instance, water (and other gases) influences surface temperature, which, in turn, affects the convection patterns \cite{lenardic2008climate, gillmann2014atmosphere}. The presence of life and interaction with water further complicates the situation. For instance, it has been suggested that life may trigger plate tectonics \cite{zhang2023photoferrotrophic}. These complex interactions across multiple levels pose challenges for modeling realistic planetary scenarios.

Future research endeavors should therefore also strive to integrate and comprehend the interconnections and feedback loops among tectonics, water circulation, landscapes, climate, and life within diverse tectonic regimes. By doing so, a more complete and detailed depiction of planetary habitability can emerge.
 
\subsection{Magnetic field generation: an illustration of the far-reaching effects of mantle dynamics}
\label{sec:mag}

The generation of a magnetic field by a terrestrial planet is a good example of the links between the interior of a planet and its surface conditions, as well as an argument in favor of considering the planet as a complete system. Firstly, magnetic fields are generated in the planetary core by a magnetic dynamo.Secondly, enough heat needs to be extracted from the core, through the mantle and out of the solid planet, in order for a dynamo to operate, again demonstrating the importance of considering a planet as a system. Thirdly, the presence of a magnetic field affects the interaction between solar radiation and the planet's atmosphere. In fact, there have been suggestions that a magnetic field may be necessary for a planet to be habitable or, at least, for life to evolve toward a more complex state \cite{cockell2016}. 

The significance of the magnetic field for supporting life is underscored by several arguments. Firstly, it provides protection to a planet's surface by shielding it from harmful solar radiation. Additionally, the magnetic field acts as a barrier against some atmospheric (especially water) loss \cite{tarduno2014}. Among the planets in our solar system, Earth is the sole planet with a self-generated magnetic field. Mars, on the other hand, bears traces of an extinct magnetic field, while in its ancient crust, which would have been active during the Noachian era, roughly 4 to 3.8 billion years ago, a period when the planet may have been less arid than it is today. 
Venus, in contrast, exhibits no evidence of an intrinsic magnetic field in recent history, nor are there any observations indicating remnants of one, which is surprising, given the usual assumption of similar structure and composition compared to Earth \cite{orourke2018}, and may indicate a radically different thermal evolution.

On Earth (and probably many rocky planets), the iron core is initially fully molten \cite{tronnes2019core} and its cooling is governed by the heat extraction by mantle dynamics. Producing a dynamo requires planetary rotation and convection in a conductive layer. Convection can occur in the liquid part of the core, triggered by the core/mantle temperature contrast or chemical differentiation. As cooling proceeds, solid iron sinks to form a solid inner core while lighter elements (such as H, C, O, Si, S, for example) rise. The motion of electrically conducting iron in an existing magnetic field induces currents generating a new magnetic field.

In both cases, magnetic fields are an expression of the thermal history and internal structure/composition of terrestrial planets. The rate of heat extraction is required to be large enough for convection to take place. On Earth, it is believed that a minimum heat flow out of the core of 6.9 TW \cite{labrosse2015} is required , which is nearly within error of Earth’s estimated CMB heat flow of 7–17 TW \cite{nimmo2015}. If Venus had an Earth-like core, the critical heat flow value would probably be around 5 TW \cite{lay2008}. However, different core properties are poorly constrained, crucially the thermal conductivity, \cite{orourke2018}.

In general, stagnant lid regimes are less effective at extracting heat from a planet's interior compared to mobile lid regimes. Consequently, a mobile lid regime, such as plate tectonics, is more likely to sustain long-lasting convection in the core. This disparity may be exemplified by the contrasting situations of Earth and Mars \cite{dehant2019geoscience}, where Mars has not generated a magnetic field for approximately 4 billion years, due to its still liquid core \cite{irving2023first} and the lack of efficient heat extraction after the primitive period. On Venus, other reasons have been proposed to be behind the absence of a magnetic field. These include an already fully solidified core \cite{dumoulin2017} or a stably stratified core \cite{orourke2018,jacobson2017} resistant to convective motion (possibly due to the absence of remixing caused by a giant impact, such as the Earth is thought to have had leading to the Moon formation).

However, recent observation of the atmospheric escape rates of O$^+$ ion on Mars, Venus and the Earth suggest that they are rather similar. 
Loss rates ranges from 2-4 $\times$ 10$^{24}$ O$^+$ s$^{-1}$ for Mars, to 1-4 $\times$ 10$^{24}$ O$^+$ s$^{-1}$ for Venus and 5 $\times$ 10$^{24}$ O$^+$ s$^{-1}$ to 5 $\times$ 10$^{25}$ O$^+$ s$^{-1}$ for Earth \cite[see][and references therin]{ramstad2021}. This implies that perhaps the magnetic field is less important than previously thought regarding atmospheric escape.

It has been suggested that a magnetic field can extend much further away from the surface of the planet than the atmosphere itself, which causes a magnetized planet to intercept a larger portion of stellar energy when compared to an unmagnetized one \cite[]{brain2016}. That additional energy would power inflated escape rates, possibly higher than for unmagnetized planets \cite[e.g.]{gunell2018}. This postulate has been challenged by \cite{tarduno2014}, who point out poorly constrained return fluxes and the variability of escape rates with the solar flux and geodynamo even for a single planet \cite{ramstad2021}. 

An undeniable effect of self-generated magnetic fields is that they modify the preferential escape location: ions flow along magnetic field lines under the effect of the solar EUV flux. With an induced magnetic field, ions could be lost in the magnetotail, as magnetic lines are open. With self-generated magnetic field, magnetic limes in the tail reconnect to the planet, preventing ions escape. However, the cusps of the magnetic field contain open field lines, and allow solar radiation to excite ions deep in the atmosphere \cite{dong2019} and accelerate them downtail through the lobes of the field, leading to their escape. This was further illustrated by models of magnetized planets with a high obliquity.

It is possible that what differs between magnetized and unmagnetized planets is the response of magnetospheres/ionospheres to intense events, like Coronal Mass Ejections (CME), which are far more numerous during planetary early evolution \cite{airapetian2020impact}. To summarize, current research casts doubt onto the classical view of the magnetic field as a shield against escape, but highlights the complexity of its interactions with solar radiation.  
The net result for terrestrial planets remains uncertain.

\section{The cessation of habitability}
\label{sec:cess}

\subsection{Liquid water and carbon cycling}

The breakdown in one of the many processes that enable long-term habitable conditions on Earth can drive the cessation of habitability. Liquid water, an energy source, CHONPS, transition metals and relevant geochemical conditions enable life to originate and evolve. The sustenance of a core-dynamo-driven magnetic field protects the Earth's biosphere from cosmic rays and charged solar particles. The steady interior evolution and plate tectonics drive outgassing and recycling of volatiles on Earth. The presence of oceans, the hydrologic cycle and the carbonate-silicate cycle (inorganic carbon cycle) allow for temperate climates by regulating CO$_2$, a vital greenhouse gas. Among these factors, the presence of liquid water on the surface and the geochemical cycling of CO$_2$ are key in ensuring habitable conditions on billion-year timescales.

Unlike Venus, where most of the near-surface CO$_2$ is present in the atmosphere, on present-day Earth, more than 99\% of near-surface CO$_2$ is expected to be locked up in the form of carbonates, courtesy of the carbonate-silicate cycle \cite[e.g. ][]{urey1952,DonahuePollack1983,wedepohl1995composition,Lecuyer2000,hartmann2012geochemical}. A number of steps are involved in recycling CO$_2$ between the atmosphere and the interior of Earth \cite{berner1983,sleep2001,krissansen-totton2018}. The continental weathering of silicate rocks \cite{walker1981,kump2000} and subsequent seafloor precipitation of carbonates locks up CO$_2$ in solids \cite{zeebe2003}. Eventual subduction of carbonates removes CO$_2$ from the atmosphere-ocean system \cite{sleep2001}. In the upper mantle, due to metamorphism, carbonates break down into silicates and release CO$_2$ back into the atmosphere \cite{kelemen2015}. 

To sustain CO$_2$ cycling, several factors are expected to be crucial, including the hydrologic cycle, fresh silicate rocks, and a planetary orbit within the habitable zone \cite{catling2017}. The incident stellar radiation (instellation) on Earth has gradually increased over the past 4.5 billion years because of the Sun's evolution on the main sequence of the Hertzprung-Russell diagram. The faint young Sun paradox suggests that early Earth must have been frozen to maintain an energy balance between solar irradiation and outgoing longwave radiation for an early Earth with similar albedo and atmospheric greenhouse effects as modern Earth  \cite{sagan1972}. The contrary evidence of an active hydrologic cycle during the Archean calls for a mechanism to ensure temperate climates during the Earth's history \cite{catling2020}. The carbonate-silicate cycle-driven steady decline in CO$_2$ partial pressure from approximately 0.1~bar (uncertainty range: 0.01--1~bar) during the Archean to a pre-industrial value of 0.28~mbar has been suggested to be a solution for the faint young Sun paradox \cite{walker1981,catling2020}. During the Phanerozoic, the biosphere has further enhanced the efficiency of CO$_2$ cycling by adding an organic component to the carbonate-silicate cycle \cite{bergman2004,hoening2020}. Which factors can break down the carbonate-silicate cycle? This question is key because it invokes a number of processes of habitability cessation that are directly or indirectly involved in the carbonate-silicate cycle. 

\subsection{Silicate weathering and carbonate precipitation}

If silicate weathering were not operational on Earth, the partial pressure of atmospheric CO$_2$ accumulated due to volcanism would already become 1 bar in 20 Myr \cite{walker1981}. This would rapidly increase the surface temperature due to the greenhouse effect of CO$_2$. Fortunately, that is not the case on Earth. The balance between weathering and outgassing on a shorter timescale of 0.1--1 Myr ensures a quasi-steady state in atmospheric CO$_2$ \cite{colbourn2015}. On a longer timescale of 0.1--1 Gyr, all planetary CO$_2$ reservoirs can attain equilibrium and prevent a runaway greenhouse state \cite{sleep2001,foley2015}. The negative feedback of silicate weathering ensures that the accumulation of CO$_2$ due to outgassing will be balanced by the loss of CO$_2$ due to silicate weathering, thereby maintaining moderate amounts of steady-state CO$_2$ on shorter timescales. On longer timescales, the intensity of silicate weathering must gradually increase with stellar luminosity.

The intensity of weathering is limited by the availability of weathering reactants, i.e., water, silicate rocks and CO$_2$ \cite{walker1981}. This leads to three limits of weathering \cite{kump2000,west2005,maher2014}. If fresh silicate rocks cannot be supplied by plate tectonics, weathering becomes rock supply-limited $W_{\rm su}$ \cite{west2005,foley2015}. However, plate tectonics may not be necessary for carbonate burial and recycling \cite[e.g., CO$_2$ cycling for stagnant-lid planets][]{foley2018,hoening2019}. From the perspective of the limited transport of hydrologically important flowpaths to fresh minerals leading to dilute water streams, supply-limited weathering is also called transport-limited \cite{kump2000,krissansen-totton2017}. Weathering becomes kinetically limited $W_{\rm ki}$ when the time available for weathering reactions is shorter than the reaction equilibrium timescale \cite{walker1981,kump2000}. If there is limited availability of water (e.g., arid climates), weathering becomes thermodynamically limited $W_{\rm th}$ \cite{maher2014,winnick2018,hakim2021}. Contrary to the kinetic limit, at the thermodynamic limit, weathering decreases as a function of temperature, resulting in a positive weathering feedback, whose impact on climate stability is not understood yet \cite{hakim2021}. To accurately model CO$_2$ cycling of temperate planets beyond modern-Earth conditions, \cite{graham2020,hakim2021} demonstrate the need to use a multi-regime weathering equation \cite[][adapted formulation]{maher2014}:
\begin{equation}
    W_{\rm total} = \left(\frac{1}{W_{\rm th}} + \frac{1}{ (\frac{1}{W_{\rm ki}} +\frac{1}{W_{\rm su}})^{-1}} \right)^{-1}.
\end{equation}
With such a formulation, the \cite{walker1981} kinetic weathering regime is recovered when $W_{\rm ki} < W_{\rm su}$ and $W_{\rm ki} < W_{\rm th}$, which is largely the case for modern Earth. The weathering is rock-supply limited when $W_{\rm su} < W_{\rm ki}$ and $W_{\rm su} < W_{\rm th}$, and the weathering is at the thermodynamic limit when $W_{\rm th} < (1/W_{\rm ki} +1/W_{\rm su})^{-1}$.

The type of silicate rocks also affects the intensity of weathering. For example, ultramafic and mafic rocks contribute up to 10 times more to weathering than felsic rocks \cite{ibarra2016,hakim2021}. Rocky exoplanets are expected to show a large diversity in silicate rock types, as studies using stellar elemental abundances suggest \cite{putirka2019,spaargaren2023}. 
Other planets may feature a wide range of gas/surface chemical reactions, that may or may not be conducive to habitability. For example, it is thought that oxidation reactions on Venus could consume or buffer oxidized components of the atmosphere, such as H$_2$O, SO$_2$ or CO$_2$ \cite{zolotov2018,gillmann2022}.

As the products of silicate weathering reactions are transferred to oceans by rivers, carbonates precipitate and settle on the seafloor for eventual subduction into the mantle. Carbonate precipitation is the following step after silicate weathering. On present-day Earth, the precipitation of calcium carbonates (calcite, aragonite) occurs almost instantaneously compared to silicate weathering \cite{sleep2001}. The precipitation of carbonates is generally neglected in carbon cycle box models because it is not the rate-limiting state among the two. However, even when continental silicate weathering takes place but the oceans are too deep, carbonate precipitation cannot occur. Below a certain ocean depth, known as the carbonate compensation depth (CCD), carbonates are unstable due to high pressures and dissolve back into the ocean \cite{broecker1987,zeebe2003}. To ensure the availability of carbonates for subduction, in principle, the CCD should be deeper than the ocean depth, which is the case for modern Earth. The CCDs for the carbonates of calcium, magnesium and iron become deeper at a higher $P_{\rm CO_2}$ and a higher temperature, thereby favouring CO$_2$ cycling for diverse carbonates \cite{hakim2023}. In the absence of continents, seafloor weathering and CO$_2$ dissolution on ocean worlds can also stabilise the long-term climate \cite{coogan2013,krissansen-totton2017,kite2018}. The total liquid water inventory and topography of terrestrial planets (which depend on outgassing and mantle dynamics) can strongly affect the drawdown of CO$_2$.

\begin{figure}[!ht]
     \centering
     \includegraphics[width=0.75\textwidth]{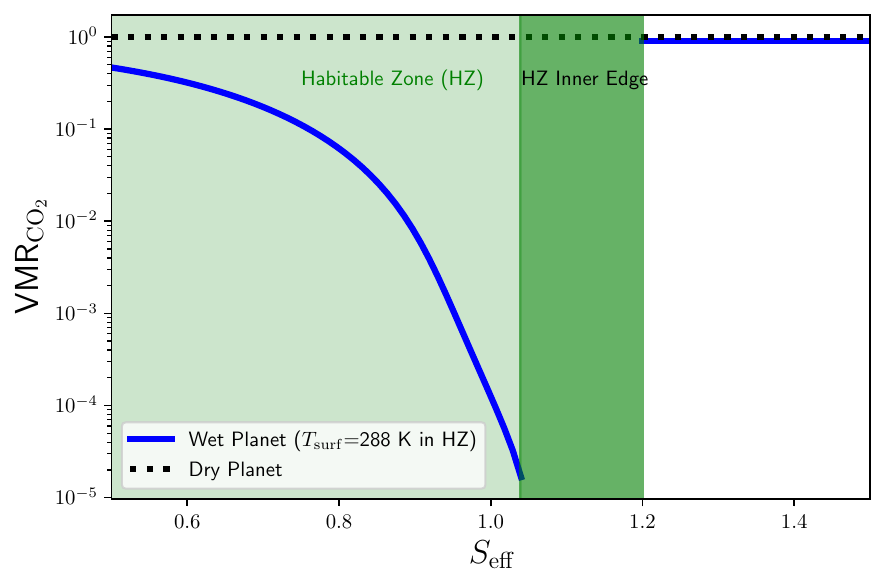}
     \caption{CO$_2$ volume mixing ratio (VMR$_{\rm CO_2}$) of two exoplanet types as a function of instellation normalised to that of modern Earth ($S_{\rm eff}$ =1). A wet planet, i.e., an Earth-like planet with surface water oceans, carbonate-silicate cycling and an N$_2$-dominated atmosphere can potentially maintain a constant surface temperature $T_{\rm surf} = 288$~K by decreasing VMR$_{\rm CO_2}$ in response to an increase in $S_{\rm eff}$ \cite[calculated using the climate model of][]{kadoya2019}. The inner HZ edge is shown as an extended zone because 1D and 3D climate models predict the inner edge to be between $S_{\rm eff}$ = 1.05--1.2 \cite{kopparapu2013,yang2014,wolf2014,wolf2015,way2018}. At higher installations beyond the inner HZ edge, water evaporation/loss followed by near-surface CO$_2$ desiccation will make the atmosphere of the wet planet CO$_2$-dominated exhibit a jump in VMR$_{\rm CO_2}$ \cite{bean2017,turbet2019,graham2020}. For a CO$_2$-dominated dry planet, the inner edge of HZ is not expected to affect VMR$_{\rm CO_2}$.} 
     \label{fig:PCO2_observability}
\end{figure}

\subsection{Observational tests}

Inside the habitable zone (HZ) for a wet planet, the carbonate-silicate cycle can be assumed to result in a constant surface temperature (e.g., $T_{\rm surf} = 288$~K, Figure~\ref{fig:PCO2_observability}). Such a stable temperature can be sustained by ensuring a higher $P_{\rm CO_2}$ at lower instellation due to a lower intensity of silicate weathering and a lower $P_{\rm CO_2}$ at higher instellation due to a higher intensity of silicate weathering. At sufficiently high instellations \cite[higher than 1.05--1.2 $S_{\rm eff}$, where $S_{\rm eff} = 1$ denotes modern Earth insolation,][]{kopparapu2013,yang2014,wolf2014,wolf2015,way2018}, the weathering feedback cannot mitigate climate warming (above a fixed energy deposition limit). This leads to the evaporation of water and an increase in the atmospheric water vapour abundance beyond the regular water vapour positive feedback \cite{catling2017}. The intricate balance between climate feedbacks and geochemical cycling is key to avoiding such an irreversible increase in surface temperature by runaway greenhouse due to the water vapour feedback mechanism, or by an additional greenhouse gas that pushes
the surface temperature beyond the critical point of water on its own \cite[the Simpson-Nakajima and Komabayasi-Ingersoll limits, respectively][]{simpson1929,komabayasi1967,ingersoll1969,nakajima1992}. The orbital distance at which the runaway greenhouse occurs is thought to define the inner HZ edge \cite{kasting1993,kopparapu2013}. A bimodal distribution in the CO$_2$ volume mixing ratio (VMR$_{\rm CO_2}$) across the inner HZ edge for wet rocky planets is statistically observable \cite{bean2017,graham2020,hoening2021}. \cite{bean2017} suggested that 20 planets could be sufficient to identify the VMR$_{\rm CO_2}$ jump at the inner HZ edge (Figure~\ref{fig:PCO2_observability}). On the other hand, \cite{lehmer2020} made statistical predictions for the planets inside the HZ. If the probability of observed planets matches a log-uniform distribution of $P_{\rm CO_2}$ regulated by the carbonate-silicate cycle, future telescopes need to observe at least 83 HZ planets \cite{lehmer2020}. \cite{schlecker2023} suggest that a high-precision transit photometry survey, such as ESA PLATO \cite{rauer2014}, can observationally identify the inner HZ edge VMR$_{\rm CO_2}$ jump for a sample of 100 exoplanets with at least 10\% exhibiting runaway climates.

It is important to note that an alternative climate state with CO$_2$ as clathrate hydrate or liquid CO$_2$ can also be stable for billions of years \cite{graham2022}. The critical absorbed flux can be crossed due to an increase in stellar radiation or changes in albedo, for instance. The moist atmosphere has a limit to how much energy it can radiate into space, which will further increase the surface temperature until the atmosphere radiates in the near-infrared. It is possible that a runaway greenhouse could lead to partial melt of the surface for large enough surface liquid water inventory \cite[about 1400 K for a few tens of bar,][]{zahnle2007}. This positive feedback can persist with almost no bounds leading to a runaway greenhouse. Venus is suspected to have undergone such a runaway greenhouse during its past \cite{DonahuePollack1983,goldblatt2012runaway}. Even on Earth, such a runaway greenhouse is expected to occur in the future as the Sun becomes brighter on its main sequence evolution.

\section{Current and future missions}
\subsection{Solar System missions}
\label{subsec:SSM}

To this date, no space mission has ever found definitive evidence for life in the solar system, beyond Earth, but habitable environments have been suggested for specific locations and times. Ancient Mars, ancient Venus, and outer solar system icy satellites are all interesting targets. Direct observation of habitable conditions is difficult, and space missions tend to investigate concomitant properties of the planet and environment, to help understand planetary evolution. 

\begin{itemize}
\item \textbf{Mars} is the most explored planet in the Solar System, apart from Earth. After negative results of the Viking mission \cite{klein1979}, it was recognized that present-day Mars is not favorable to life, despite minute signs that habitability might not be far off martian conditions. Any habitable phase would have occurred more than 3.8 Gyr ago.

Further exploration operated under the "follow the water" guideline and attempted to characterize the ancient geology of Mars and put time constraints on water-related, geological and mineralogical features \cite[e.g.][]{carr2007,bibring2006,squyres2005}. 
Mars, once, had more water (liquid, at times) than at present, even if it remains uncertain whether habitable conditions were stable or episodic, or if the climate was warm or cold \cite[see][for instance]{wordsworth2016,bishop2018}.

Further investigation has targeted interactions between the interior and the atmosphere. Phoenix (which arrived on Mars in 2008) has investigated polar region rhegolith and found that it has interacted little with liquid water in recent history \cite{kounaves2014}. The Curiosity rover (landed in 2012) and Perseverance (landed in 2018) looked into the possible subsurface origin of methane \cite{webster2015,hu2016}, and provided data regarding surface mineralogy, past traces of water, the lower atmosphere and geology.Finally, Tianwen-1 (China National Space Administration, CNSA) landed in 2021 and notably explored the relation between surface features, the past presence of water and the nature of the martian surface.

Insight (2018-2022) aimed to cover the deep interior of Mars and its structure, composition and dynamics, in particular using a seismometer \cite{banerdt2020,irving2023first}. The mission recorded hundreds of Mars-quakes, and constrained the size of the core and its fraction of light elements. 

Future Mars-related astrobiology missions include multiple plans for a Mars Sample Return (NASA), as many questions require direct analyses of martian material \cite{mattingly2011,beaty2019}. Plans for the Tianwen-3 mission (CNSA) also include sample return (launch scheduled for 2028). Finally, the two-part Exomars mission (2016 and 2028 launches) attempts to investigate biosignatures, water distribution, carbon isotopes and their possible biologic origins. 
\\

\item \textbf{Venus} has been relatively little explored compared to other targets in the solar system.

Early remote observation, Mariner 2 and the soviet Vega and Venera landers (in the 60's and the 70's) revealed the crushing 92 bar surface pressure, and $\approx$740 K surface temperature caused by its massive CO$_2$ atmosphere \cite[see][for a detailed overview]{o2022venus}. 

Ancient habitable Venus has nonetheless been suggested \cite{way2020venusian}, but is debated \cite[see][]{gillmann2022,westall2023} and depends on very specific atmospheric conditions \cite{Turbet2021}. 
Present-day habitability of the atmosphere near the cloud layer has been suggested \cite{limaye2021venus} but remains unconfirmed, while recent claims of the observation of phosphine \cite{Grea21} in the atmosphere have again triggered heated debate \cite{Trom21,Encr20,Snel20,Thom21,Vill21}. 
Beyond invaluable but imprecise in-situ measurements (Vega, Venera), Magellan (1990) and Venus Express (2006) have been the source of most of the current scientific knowledge of the surface and atmosphere of Venus \cite{o2022venus}. Additional observation was performed by Akatsuki (2015) and the Parker Solar Probe (2018), as well as other missions en-route toward other targets \cite{taylor2018venus}.

There is a general consensus that we still know very little about Venus' interior structure and composition, and lower atmosphere. Yet, the evolution of Venus' interior and how it may have caused a divergent evolution compared to Earth are critical research avenues for habitability. 

This caused a resurgence of interest in Venus with several high-profile missions planned in the next decades: DaVinci and VERITAS (US), EnVision (EU), VENERA-D (RU), Shukrayaan-1 (India), VOICE (China). 

All six missions are designed to investigate the past of the planet and the processes that affect the evolution of its surface conditions and volcanic processes. 
Ancient traces of water are also investigated, throughout the atmosphere, and near possibly ancient material \cite[the tesserae][]{byrne2021venus,whitten2021venus,smrekar2018venus}. The main goal of those missions is not just to know if Venus was habitable, but to understand why it is not at present-day, so that this knowledge can be applied more broadly \cite{way2023synergies,gillmann2022}.
\\

\item \textbf{The icy moons of Jupiter and Saturn} are considered important targets for astrobiology missions. The Cassini Huygens mission opened the door to further exploration of the satellites of giant outer solar system planets. The landing on Titan and discovery of its surface and liquid hydrocarbon lakes has seen suggestion of methanogenic life \cite{mckay2005}. Io, Enceladus, Ganymede and Europa are all suspected of harboring (possibly habitable) interior liquid water oceans beneath their icy surface \cite{chyba2001,barge2021,vance2014,sephton2018}. Future missions include ESA's Juice toward Europa, Ganymede and Calisto \cite{grasset2013}. Europa is also the target of NASA's Europa Clipper \cite{phillips2014}. NASA's second mission, Dragonfly \cite{lorenz2018,barnes2021} would place a rotorcraft on Titan to study biosignatures and chemistry. 
\end{itemize}

\subsection{Exoplanet missions}
In the context of detecting and, even more importantly,  characterizing the (atmospheric) properties of temperate, terrestrial exoplanets it is important to consider a few aspects: 

\begin{itemize}
    \item {\bf M-type stars vs. Solar-type stars:} Depending on the effective temperature and luminosity of the host star\footnote{Note that we are primarily concerned with main-sequence stars here.}, the orbital period (and hence mean orbital separation) of habitable zone planets varies significantly, ranging from up to a few hundred days for stars similar or slightly hotter than the Sun, to less than a few tens of days for cool, low-luminosity M-type dwarf stars. This has important implications, which techniques can be used for atmospheric investigations of the objects. 
    
    \item {\bf Transit spectroscopy and secondary eclipse observations vs. direct detection methods:} Currently, the most successful approach to observe the atmospheres of exoplanets are transit spectroscopy and secondary eclipse measurements \cite[for an introduction, see,][]{seager2010}. These techniques have a strong bias towards large, close-in planets. And while the James Webb Space Telescope (JWST) and also ESA's Ariel mission \cite[][]{tinetti2018} will revolutionize our understanding of exoplanet atmospheres for hot/warm gas dominated planets, they will not be able to probe a large sample of rocky, temperate exoplanets. In the most favorable cases, which are small planets transiting very low-mass stars \cite[such as the well-known Trappist-1 system,][]{gillon2017}, JWST may be able to tell if these objects have an atmosphere at all, which will be an important milestone in exoplanet science \cite[for first results, see,][]{greene2023,zieba2023}, but an in-depth characterization and comprehensive inventory of atmospheric constituents will likely be out of scope \cite[e.g.,][]{krissansen-totton2018,lustig-yaeger2019}. The reason is very simple: despite the large 6.5-m primary mirror, the favorable planet-star brightness ratio, the proximity of the Trappist-1 system of only $\approx$12\,pc, and the ability to observe and stack several transits due to the very short orbital periods of the planets, the signal-to-noise of the data will remain limited. This is the reason why, in the future, a main focus will be on direct detection methods that are largely independent from orbital properties and do not require an exoplanet to transit its host star\footnote{For completeness we mention here the ambitious and inspiring \emph{Nautilus} project that seeks to search for biosignatures in the atmospheres of hundreds of transiting terrestrial exoplanets with a fleet of low-cost spacecraft \cite[][]{apai2019AJ}.}; the vast majority of exoplanets does -- in fact -- not transit in front of their stars!  

    \item {\bf Reflected light vs. Thermal emission:} Accepting that direct detection techniques are a promising way to probe many of the temperate, terrestrial exoplanets within 10-25 pc from the Sun, one needs to realize that there are two ways to directly detect light from these objects: one can either probe the starlight reflected by the planets (this is typically done at UV, optical, and near-infrared wavelengths between $\sim$0.2 and $\sim$2.5$\;\mu$m) or one can probe the planets' intrinsic thermal emission (which, for temperate  objects similar to the terrestrial planets in the Solar System, can be done at mid-infrared wavelengths between $\sim$3$\;\mu$m and $\sim$20$\;\mu$m).
\end{itemize}

With currently available telescopes (on ground and in space), we cannot directly detect small, habitable zone exoplanets. It's technically not feasible as none of the existing facilities or instruments provides the full combination of required spatial resolution, contrast performance, and sensitivity. For the required contrast, we remind the reader that in reflected light at optical wavelengths the Earth is a factor of $\approx 10^{10}$ fainter than the Sun, and at a wavelengths of $\sim$11$\;\mu$m, where the thermal emission of Earth peaks, the flux difference still amounts to $\approx 10^{7}$ \cite[e.g.,][]{desmarais2002}. For planets orbiting low-luminosity M-dwarfs these values are typically a factor of 100 lower. 

Considering the points listed above, it becomes clear that a number of new instruments and missions will be needed for a comprehensive investigation of a large number of temperate terrestrial exoplanets orbiting various kinds of host stars and across a large wavelength range:

\begin{itemize}
    \item Probing nearby exoplanets orbiting low-luminosity M-dwarfs in reflected light will be the primary discovery space of optimized instruments installed at ELTs \cite[e.g.,][]{kasper2021}. With their 30-40 m primary mirrors they will provide sufficient spatial resolution to spatially separate the signal from the planet from that of its host star on the detector, and the combination of extreme adaptive optics systems with high-resolution spectrographs will deliver the required contrast performance \cite[][]{snellen2015}.
    \item Probing nearby exoplanets orbiting Sun-like stars in reflected light will required dedicated space missions. The stringent contrast requirement (see above) can likely not be achieved by ground-based instruments due to the disturbing effects from the Earth atmosphere. In fact, a large, single aperture exoplanet imaging mission was recently recommended in the context of NASA's 2020 Astrophysics Decadal Survey\footnote{{https://www.nationalacademies.org/our-work/decadal-survey-on-astronomy-and-astrophysics-2020-astro2020}}. The main science driver for this so-called \emph{Habitable World Observatory (HWO)} is to detect and investigate $\approx$25 Earth-like planets around nearby Sun-like stars in reflected light. While the exact wavelength coverage is still under investigation -- in particular the short- and long-wavelength cut-off in the UV and near-infrared will be carefully examined -- the aperture size shall be $\approx$6\,m and hence in between those of the \emph{LUVOIR} and \emph{HabEx} mission concepts, which have been studied in preparation for the Decadal Survey \cite[][]{luvoir2019,habex2020}. We remind the reader that the \emph{Hubble Space Telescope (HST)} has a primary mirror of 2.4\,m. The oxygen A-band (759–770\,nm) is the main diagnostic for \emph{HWO} to search for indications of biological activity in exoplanet atmospheres, but absorption bands of water, and potentially also carbon dioxide and methane, shall also be accessible. First studies quantifying how well Earth-like planets can be characterized by a future reflected light mission like \emph{HWO} have been carried out \cite[][]{feng2018AJ}.
    \item Finally, probing the thermal emission of nearby exoplanets orbiting any type of host star, requires a large, space-based nulling interferometer. Even with the 39\,m ELT of the European Southern Observatory the thermal background noise at mid-infrared wavelength for ground-based observations will be prohibitively high so that only very few of the nearest stars can be imaged directly in the search of temperate terrestrial exoplanets \cite[][]{quanz2015,bowens2021}. From space, however, a mission like the \emph{Large Interferometer For Exoplanets (LIFE)} \cite[][]{quanz2022a} could detect hundreds of nearby exoplanets including dozens similar to Venus and Earth in terms of size and energy influx \cite[][]{quanz2022b,kammerer2022}. An exoplanet's thermal emission allows for a robust characterization of its atmospheric properties such as pressure-temperature profile and composition \cite[][]{line2019} as it is less affected by disturbing effects such as clouds \cite[][]{kitzmann2011} compared to observations in reflected light. In the case of a \emph{LIFE-}like mission, habitable conditions on an Earth-like planet can readily be inferred from moderate resolution mid-infrared spectra throughout most of Earth's history \cite[][]{konrad2022,alei2022} and the mid-infrared wavelength range provides access to a variety of atmospheric biosignatures gases \cite[e.g.,][]{schwieterman2018}. We emphasize that the direct detection of temperate terrestrial exoplanets at mid-infrared wavelengths with the goal to identify habitable -- and potentially even inhabited -- exoplanets was given very high scientific priority in the \emph{Voyage 2050} program of the European Space Agency (ESA) and has been identified as a potential topic for a future L-class mission in ESA's science program\footnote{https://www.cosmos.esa.int/web/voyage-2050}. 
\end{itemize}
Looking at the list above one can be hopeful that in the coming 20-30 years we will indeed have a suite of instruments and missions that strongly complement each other in the search for habitable worlds beyond the Solar System. These projects are \emph{not} in competition with each other; in addition to the unique scientific diagnostic that each of them possesses, the combined power of reflected light and thermal emission data will provide a truly holistic picture of temperate terrestrial exoplanets orbiting a wide range of stellar types in our cosmic neighborhood.

\section{Conclusions and summary}

Habitability is a complex notion dependent on many interwoven processes. The qualitative definition of habitability being a set of conditions that are conducive to life is generally insufficient for any precise assessment or prediction. On the other hand, more quantitative definitions are strongly biased by the only example of a unquestionably habitable world we possess. 

At its core, the habitability of a planet is generally restricted to assessing the suitability of the conditions at its surface for self-sustaining biotic reactions to occur; that is, the response of a planetary atmosphere to the energy input (such as stellar radiation, for example). However, that response depends on the planet's time-integrated atmospheric structure and composition, which are governed by i) the initial availability of life-essential elements in the planetary body and ii) a myriad of interacting mechanisms and feedback processes. They can operate both inside the atmosphere (chemistry, dynamics) or outside (volatile or energy exchange with the planetary interior). The fate of water is especially important for life (as we know or imagine it, at least), and constitutes the first order guide for habitability models and detection. 

The relative importance of these different processes changes depending on the stage of planetary evolution and time. Two of the main drivers of changes are the evolution of the host star and that of the silicate planet. In this review, we have discussed some of the most high profile questions regarding the role of the interior of terrestrial planets in controlling habitability, such as the conditions for its onset, the role of interior dynamics, and how it is maintained and could cease. 

The nature of terrestrial planets, by definition, is distinct from those of gas giants that have nearly their full complement of stellar gases. By contrast, terrestrial planets are relatively `dry' \cite{aston1924}. Therefore, the masses of atmospheres around terrestrial planets are \textit{products} of the bulk composition of the planet and the prevailing pressure-temperature conditions under which the atmosphere formed and evolved through time. Given that $>$99.95 \% of the bulk silicate mass of planets resides in rocks \cite{palmeoneill2014}, the proportions of O to other rock-forming elements, namely, Fe, Mg and Si, control the nature of atmospheres around rocky planets through the oxygen fugacity, \textit{f}O$_2$. 

As such, a planet's interior governs the long-term evolution of the atmosphere by exchanging mass (in the form of volatile species) and heat with it. The net flux of outgassing (e.g., volcanic exhalation) and ingassing (burial of volatile elements into the mantle and crust) defines the bulk composition of the atmosphere (and surface) and thus its water inventory \cite[see][]{marty2012,hirschmann2018}. We have underlined the importance of the magma ocean stage in setting the initial conditions that can lead to habitable surface conditions (section \ref{sec:onset}). Later, habitability is maintained or fails due to a delicate balance in volatile exchange. On the Earth, the relevant species are water and CO$_2$ owing to the relatively oxidised nature of its mantle \cite[$\Delta$IW+3.5 at present in the upper mantle,][]{frostmccammon2008}, but on other planets could contain other gaseous species, namely H$_2$ or CO on more reduced planets, and SO$_2$ on more oxidising planets.

This field of investigation is still in its infancy, as results of the past few decades have highlighted the vast complexity of the interactions leading to volatile exchange \cite[e.g.,][]{fegleyschaefer2014,hirschmann2018,Zahnle_2020,grewal2020speciation,bower_etal2022}. They involve all layers of the planet, from the core to the upper atmosphere, in addition to the more obvious atmosphere/mantle interface. This is an inherently multidisciplinary endeavour that requires a deep understanding of all the parts of the volatile cycle 
In short, understanding habitability requires us to understand planetary evolution as a whole, an equally exciting and daunting prospect.

Current scientific efforts include many missions especially focused on the theme of habitability, either in our Solar System or toward exoplanets. It is tempting to look for the safest targets, selecting objects similar to Earth, but it is better to cast a wide net and see what comes up. Many missions attempt this with planets that have very different conditions from Earth, such as Venus, for example, to understand the mechanisms of planetary evolution, or for exoplanets, to build-up statistics beyond our small solar system.
After all, there is no guarantee that Earth is the unique blueprint for habitability. At present, we still understand the topic far too little to make solid predictions. Instead, we would reap tremendous benefits form observation of the diversity of possible planetary evolution outcomes.

\section*{Acknowledgements}
Funding: Parts of this work has been carried out within the framework of the National Centre of Competence in Research PlanetS supported by the Swiss National Science Foundation (SNSF) under grants 51NF40\_182901 and 51NF40\_205606. PAS thanks the SNSF via an Eccellenza Professorship (203668) and the Swiss State Secretariat for Education, Research and Innovation (SERI) under contract number MB22.00033, a SERI-funded ERC Starting Grant '2ATMO'. KH acknowledges the FED-tWIN research program STELLA (Prf-2021-022), funded by the Belgian Science Policy Office (BELSPO). Author contributions: CG directed this review, organized the paper, wrote sections 1 and 6 and contributed to sections 2,3 and 5. KH was in charge of section 4. DL was in charge of section 3. SQ was in charge of section 5. PS was in charge of section 2. All authors discussed, commented and revised the paper. Competing interests: The authors declare that they have no competing interests.

\printbibliography

\end{document}